\documentclass[preprint,3p,times,preprint,showpacs,amsmath,superscriptaddress,floatfix]{elsarticle}

\usepackage{amssymb}
\usepackage{graphics}
\usepackage{float}
\usepackage{color}
\usepackage{epsfig}
\usepackage{epstopdf}
\usepackage{url}

\usepackage{graphicx}
\usepackage{dcolumn}
\usepackage{multirow}
\usepackage{color}
\usepackage[normalem]{ulem}
\newcounter{bla}

\journal{Computer Physics Communications}

\begin{document}

\begin{frontmatter}


\hyphenpenalty=5000

\tolerance=1000

\title{TNSPackage: A Fortran2003 library designed for tensor network state methods}

\author[add1,add2]{Shao-Jun ~Dong}

\author[add1,add2]{Wen-Yuan ~Liu}

\author[add1,add2]{Chao Wang}

\author[add1,add2]{Yongjian ~Han\corref{cor1}}
\ead{smhan@ustc.edu.cn}

\author[add1,add2]{G-C ~Guo}

\author[add1,add2]{Lixin He\corref{cor1}}
\ead{helx@ustc.edu.cn}

\cortext[cor1]{Corresponding authors}

\address[add1]{CAS Key Laboratory of Quantum Information, University of Science and Technology of China, Hefei, Anhui, 230026, China}
\address[add2]{Synergetic Innovation Center of Quantum Information and Quantum Physics, University of Science and Technology of China, Hefei, Anhui, 230026, China}

\begin{abstract}
{Recently, the tensor network states (TNS) methods have proven to be very powerful tools to
investigate the strongly correlated many-particle physics in one and two dimensions.
The implementation of TNS methods depends heavily on the operations of tensors,
including contraction, permutation, reshaping tensors, SVD and son on.
Unfortunately, the most popular computer languages for scientific computation,
such as Fortran and C/C++ do not have a standard library for such operations, and therefore make the coding of TNS very tedious.
We develop a Fortran2003 package that includes all kinds of basic tensor operations
designed for TNS. It is user-friendly and flexible for different forms of TNS, and therefore
greatly simplifies the coding work for the TNS methods. }
\end{abstract}


\end{frontmatter}

% \linenumbers

% Computer program descriptions should contain the following
% PROGRAM SUMMARY.

  %Delete as appropriate.

\begin{small}
\noindent
{\bf PROGRAM SUMMARY}

\noindent
{\em Program Title: TNSP}                                        \\
{\em Licensing provisions}: GNU General Public License, version 3\\
%  CC0 1.0/CC By 4.0/MIT/Apache-2.0/BSD 3-clause/BSD 2-clause/GPLv3/CC BY NC 3.0 }                                   \\
{\em Programming language:} Fortran2003                                   \\
{\em External routines:} BLAS, LAPACK, ARPACK\\
{\em Operating system:} Linux, MacOSX \\
{\em Nature of problem:} \\
The implementation of Tensor Network State (TNS) methods depends heavily on the operations of tensors.
Unfortunately, the most popular computer languages for scientific computation,
such as Fortran and C/C++ do not have a standard library for such operations, and therefore make the coding of TNS very tedious. \\
{\em Solution method}:
We develop a Fortran2003 package that includes all kinds of basic tensor operations
designed for TNS, which greatly simplifies the coding work for the TNS methods. \\
  %Describe the method solution here.
{\em Additional comments including Restrictions and Unusual features}:\\
A gcc-4.8.4 or later version is required to compile the code.\\

\end{small}

\section{Introduction}
\label{sec:introduction}

One of the biggest challenges in modern condensed matter physics is to develop efficient numerical methods to
solve the strongly correlated many-particle physics. As for strongly interacting systems, where conventional perturbation theory fails, numerical simulation plays a crucial role to reveal the nature of quantum many-body physics. Several popular methods, including the exact diagonalization (ED), Quantum Monte Carlo(QMC) methods \cite{Kashurnikov2007,Sandvik2002} and the density matrix renormalization group (DMRG) method \cite{white92,white93} , have been widely used and achieve great success. However, there are some limitations for the previous methods: ED methods suffers the ``Exponential Wall'' problem and the QMC suffers the notorious sign problem when simulating frustrated systems and fermion systems~\cite{Troyer05}, whereas DMRG is limited to 1D or quasi-1D systems and does not work well for higher dimension systems\cite{Schollwoeck2011}. It is pressing to develop new efficient numerical algorithms.

Recently, tensor network states (TNS), including matrix product states (MPS)\cite{Vidal2003}, projected entangled pair states (PEPS)\cite{Verstraete2004} etc., are proposed to describe many-body physics inspired by the quantum entanglement theory. The TNS methods provide a promising scheme to investigate the systems that are not tractable by the previous methods. In the case of 1D, matrix product states (MPS) which constitute the variational space of DMRG, has been established as the leading method for the simulation of the statics and dynamics of one-dimensional strongly correlated quantum systems~\cite{Karrasch2012,Huang2014}, both at zero~\cite{Vidal2007} and finite temperatures~\citep{Barthel2013}. As a natural extension of MPS to higher dimensions, PEPS show great potential to solve some long-standing problems\cite{Verstraete06,Verstraete2008,wang16,liu16}. Besides, TNS supply great flexibility for different problems, such as projected entangled simplex states (PESS)\cite{xie2014} and multi-scale entanglement renormalization ansatz (MERA) \cite{Vidal2008} which have been successfully applied to simulate frustrated magnets\cite{xiang2016,Evenbly2010}.

In the TNS scheme, there are many differences for programming codes when adopting different TNS forms including MPS, PEPS, PESS, MERA, string-bond states (SBS) ~\cite{Cirac2010} and so on. Therefore unlike the first-principles packages based on density functional theory, it is impossible to develop a universal code that includes all these methods.
However, the TNS methods do have very similar features, where they all depend heavily on some basic
operations of high rank tensors, even though with different TNS forms. Unfortunately,
so far there is no standard library for such tensor operations in the most popular computer languages for scientific computations, e.g.,
Fortran and C/C++.
Directly using the high dimensional arrays, makes the codes tedious, fragile, low efficient and hard to maintain.

To solve this problem, we develop a Fortran2003 package that integrates some most used tensor operations,
including multiplication, contraction, permutation, singular value decomposition (SVD) and other matrix decomposition operations, etc..
The package is very flexible, supporting all kinds of tensor formats, and data types, including
integer, single or double precision real number, single or double precision complex number, logical and character.
Some high performance mathematic packages, including Arpack \cite{Arpack}, Lapack \cite{LapackAndBlas} and Blas \cite{LapackAndBlas}
have been adopted to speed up the performance in the package.
We extensively use advanced language features of Fortran2003, such as an object-oriented programming
style, to improve the readability and re-usability of the package.
We have successfully implemented the SBS method~\cite{Dong,WangChao} and PEPS methods \cite{liu16} for different many-particle Hamiltonian and geometries
using the package, which greatly reduces the workload for the coding.

The rest of the paper is organized as follows.
We first introduce the basic theory of TNS in Sec. \ref{sec:theory} and their applications in
many-body physics in Sec.~\ref{sec:theory2}.
We summarize some most often used basic operations in TNS methods in Sec.~\ref{sec:operations},
and then introduce the main features of the TNSpackage in Sec.~\ref{sec:overview}.
In Sec.~\ref{sec:examples}, we give an example of how to implement a TNS algorithm (here PEPS) using the TNSpackage.
We conclude in Sec.~\ref{sec:summary}.

\section{Basic theory of Tensor Network States}
\label{sec:theory}

In this section, we introduce the basic theories of TNS.
We start with the one-dimension matrix product states (MPS)\cite{Vidal2003,Verstraete2008}.

\subsection{Matrix product states(MPS)}

Consider a 1D lattice with $N$ sites and a physical degree freedom $s$ on each site. A general quantum state of this system can be expanded as
\begin{equation}
	|\Psi\rangle=\sum_{\{s\}}c_{s_1,s_2,\dots,s_N}|s_1,s_2,\dots,s_N\rangle \, ,
\end{equation}
where $c_{s_1,s_2,\dots,s_N}$ is the coefficient for the basis $|s_1,s_2,\dots,s_N\rangle$, and $\sum_{\{s\}}|c_{s_1,s_2,\dots,s_N}|^2=1$. Generally, the total number of coefficients is $2^N$ which increases exponentially with the size of the system.
To reduce the parameters of the many-body quantum state, the matrix product state (MPS) is introduced in which coefficients $c_{s_1,s_2,\dots,s_N}$ are expressed as traces of a set of matrices\cite{Schollwoeck2011,Schollwoeck2012}, i.e.,
\begin{equation}\label{MPSPsi}
|\Psi\rangle=\sum_{\{s\}}{\rm tr}(A[1]^{s_1}A[2]^{s_2}\cdots A[N]^{s_N})|s_1,s_2,\dots,s_N\rangle\, ,
\end{equation}
where $A[i]$ is a tensor corresponding to the site $i$, and the index $s_i$ is called the physical index of the tensor $A[i]$ since it corresponds to the physical degree freedom on site $i$. For each basis $|s_1,s_2,\dots,s_N\rangle$, the physical index of $A[i]$ is fixed as $s_i$ and the resulted tensor, $A[i]^{s_i}$, will be a matrix besides the endpoints of the 1D lattice (the tensor form of the endpoints depend on the boundary condition: for the open boundary, it will be a vector for fixed physical index; for the periodic boundary condition, it is a matrix). For the matrix $A[i]^{s_i}$, it has the matrix indexes $l_i$ and $r_i$ which are called virtual indices of the $A[i]$ since they are not physical and will be contracted in the state. This state is therefore named as matrix product state. The dimension of the virtual indices, denoted by $D$, is the controlling parameters of the state. The MPS with proper virtual $D$ has been used as a successful variational space in DMRG method \cite{Schollwoeck2011}. In addition, the ground state of a 1D gapped system and the slightly entangled state can be well approximated by an MPS \cite{Hastings2007}, the precision of the approximation can be well controlled by the virtual dimension $D$: the larger $D$, the better approximation.

\begin{figure} [!hbp]
	\begin{center}
		\resizebox{60mm}{!}{\includegraphics{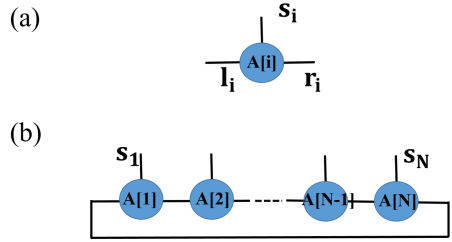}}
		\caption{(a) The graphical representation of tensor $A[i]^{s_i}_{l_{i},r_i}$ at $i$-th site. Every leg denotes a tensor index.
(b) The graphical representation of the MPS corresponding to Eq.~(\ref{MPSPsi}).
The connected legs are virtual bonds that are summed over.}\label{graphicalMPS}
	\end{center}
\end{figure}

It is very convenient to use the graphical notation~\cite{Schollwoeck2012} to denote MPS instead of the formula like Eq.~(\ref{MPSPsi}). In the graphical notation, a tensor is represented by a geometrical shape such as a circle or a diamond, and its indices are represented by outgoing legs. The legs corresponding to physical indices are called physical legs, and the legs corresponding to virtual indices are called virtual legs. As shown in Fig.~\ref{graphicalMPS}, the upper one  represents a single tensor $A[i]$. It has 3 indices, represented by 3 legs, with physical leg pointing upward and virtual legs pointing leftward and rightward. The bottom picture represents an MPS with the periodic condition. The connected legs are called bonds, which represent the virtual indices that are summed over.

From the quantum information point of view, more importantly, with deeper understanding of the relation between area law and the DMRG method \cite{Schollw}, the MPS can be interpreted in another way which can be very conveniently extended to higher dimensions:
\begin{equation}\label{MPSPsi3}
|\Psi\rangle=\sum_{\{s\},\{r\},\{l\}}A[1]^{s_1}_{r_1}|s_1\rangle\langle r_1|\  A[2]^{s_2}_{l_2r_2}|s_2\rangle\langle l_2r_2|\cdots A[N]^{s_N}_{l_N}|s_N\rangle\langle l_N|\ EPR_{12}\cdots EPR_{N-1,N} \; ,
\end{equation}
where $|l_i\rangle$ and $|r_i\rangle$ are states in a $D$ dimensional virtual Hilbert space, and
\begin{equation}
EPR_{i,i+1}=\frac 1{\sqrt D}\sum_{j=1}^D |r_i=j,l_{i+1}=j\rangle
\end{equation}
are the maximal entangled state on the bond between site $i$ and site $i+1$. The configuration of the EPR pairs automatically satisfies the area law and can be viewed as the matrix of the MPS state in the virtual space.
 For each site, there is a projector, $A[i]^{s_i}_{l_ir_i}|s_i\rangle\langle l_ir_i|$, which project the state from the virtual space into the physical space. Any set of projectors will define a quantum many-body state. The projectors then define a variational space. The quantum information theory \cite{Bennett}
tells us that the entanglement can not be increased under the projections, that is, the resulting state (MPS) also satisfies the area law. With this understanding of the MPS, it is natural to extend the MPS to the higher dimension.

\subsection{Projected entangled pair state (PEPS)}

We now generalize MPS to higher dimension, namely the projected entangled pair state (PEPS). For each 2D lattice, we assign each site $i$ ($i$ is a coordination in 2D) a tensor $A[i]^{s_i}_{d_{i,1} d_{i,2},\dots}$, where $d_{i,1} d_{i,2},\dots$ are virtual indices corresponding to virtual legs to be connected with neighboring sites. Generally, the PEPS can be written as,
\begin{equation}\label{MPSPsi4}
|\Psi\rangle=\sum_{\{s\},\{d\}}\prod_iA[i]^{s_i}_{d_{i,1} d_{i,2},\dots}
|s_i\rangle\langle d_{i,1} d_{i,2},\dots|\prod_{\langle i,j\rangle} EPR_{ij}\, ,
\end{equation}
where $EPR_{ij}$ is the EPR pair on bonds connecting site $i$ and site $j$.

This can also be expressed compactly as
\begin{equation}\label{MPSPsi2}
|\Psi\rangle=\sum_{\{s\}} \mathcal{C}\Big[\prod_iA[i]^{s_i}_{d_{i,1} d_{i,2},\dots}\Big]|{s_1,s_2,\dots,s_N}\rangle
\end{equation}
where $\mathcal{C}$ means to contracts all connected virtual legs.

\begin{figure} [!hbp]
		\begin{center}
       \includegraphics[width=0.7\linewidth]{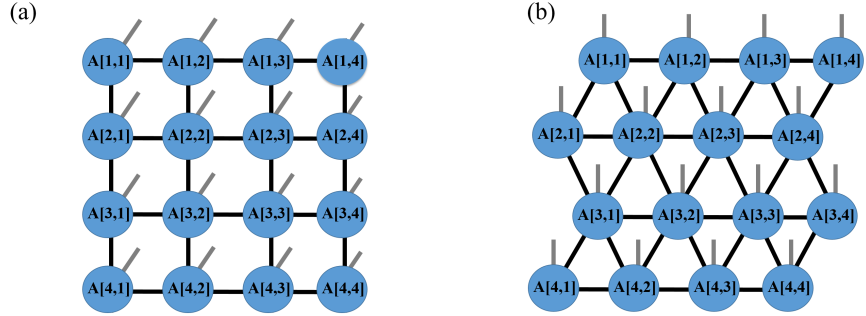}
		\caption{(a) A PEPS on a 4$\times$4 square lattice, where the tensors have maximum 4 virtual legs.
(b) A PEPS on a 4$\times$4 triangular lattice, where the tensor have 6 virtual legs. Each circle corresponds to a site in the lattice.
The free legs are the physical indices.
}\label{graphicalPEPS}
		\end{center}
\end{figure}

The number of the virtual legs can be the coordination number of the lattice, e.g., 4 for a square lattice, 3 for a honeycomb lattice and 6 for a triangular lattice.  The PEPS of a square lattice and a triangular are graphically represented in Fig.~\ref{graphicalPEPS}. For the square lattice, each tensor in the network has 4 virtual legs that connect to neighbor sites and a physical leg open, whereas for the triangular lattice, there are 6 virtual legs for each tensor.
The PEPS satisfies area law \cite{Eisert10}, therefore it is an efficient representation of the
ground states of some strongly correlated many-body systems.

\subsection{Other tensor network states(TNS)}

The MPS and PEPS both satisfy the area law (MPS in 1D, PEPS in 2D) which can be used to well approximate the ground state of many systems. Along the same way, some other TNSs which also satisfy the area law are constructed, such as SBS\cite{Cirac2010}, PESS\cite{xie2014}. The different TNS has its own advantages in certain situations.

However, there are some systems whose volume entropy is beyond the area law, such as the free fermion system \cite{Wolf}. As a result, the former states can not be used as the efficient approximation of the states in this situation. Some TNSs which are beyond the area law can also be constructed, e.g., MERA\cite{Vidal2008}.

\section{Tensor Network States in many-body physics}
\label{sec:theory2}

TNS have many applications in different physical fields, from condensed matter physics to black hole \cite{Pastawski15}. In this section, we introduce the application of TNS in many-body physics. We start from the method to use TNS to access the ground state of a quantum system.

\subsection{Use TNS to access the ground state of a quantum system}

As we have mentioned, the ground states of the quantum systems, especially for the gapped systems, can be effectively approximated by the TNS with proper virtual
bond dimension $D$. In this section, we introduce some methods to find an optimal TNS with fixed parameter $D$ to approximate the ground state for
a specific Hamiltonian. There are two major methods to optimize the TNS: (i) imaginary time evolution method and (ii) variational method.
We take the 1D MPS as an example to explain these methods, which can be generalized to other TNS.
	
\subsubsection{Imaginary time evolution method}

In the imaginary time evolution method, we start from a randomly chosen MPS  with fixed parameter $D$.
The ground state is obtained by evolving the initial state under imaginary time for long enough time, i.e.,
\begin{equation}\label{TEBDformula1}
	|0\rangle=\lim_{\tau\rightarrow\infty}e^{-\tau H}|\Psi\rangle \, ,
\end{equation}
where $|\Psi\rangle$ is the initial state, $|0\rangle$ is the ground state of the Hamiltonian. The algorithm originates from the fact that we can expand the initial state by energy eigenstates of the Hamiltonian, as $|\Psi\rangle=c_n|n\rangle$, where $|n\rangle$ is the eigenstate with $n$th smallest eigenvalue.
We have
\begin{eqnarray}\label{TEBDformula}
&e^{-\tau H}|\Psi\rangle\\ \nonumber
&=&\sum_{n=0} c_n e^{-\tau E_n}|n\rangle \\ \nonumber
&=&e^{-\tau E_0}\Big[c_0|0\rangle+\sum_{n=1}e^{-\tau (E_n-E_0)}c_n|n\rangle\Big] \propto |0\rangle \, .
\end{eqnarray}
It is easy to see only the lowest energy state $|0\rangle$ survives after imaginary time evolution for long enough time.

Generally, the operator $e^{-\tau H}$ is a global operator.
From calculation aspect, we can not deal with the whole system whose Hilbert space is huge. Therefore, we have to approximate the imaginary time evolution $e^{-\tau H}$ by a set of operators which only act on several local sites. We first divide the time evolution into $N$ segments, i.e.,
\begin{equation}
e^{-\tau H}=\big(e^{-H\Delta\tau}\big)^N \, ,
\label{eq:TE}
\end{equation}
where $\Delta\tau=\frac\tau N $ is a small time step.
We then decompose the Hamiltonian $H$ into summation of $m$ terms of local Hamiltonian,
%Generally, the Hamiltonian $H$ is composed by a set of local Hamiltonians, we divide these local Hamiltonians into $m$ terms
\begin{equation}
	H=H_1+\cdots+H_m \, ,
\end{equation}
each of which contains only commuting indivisible terms.
For a very short time ($\Delta\tau\ll 1$), the time evolution can be well approximated by the production of a series of operators using (2nd order)
Trotter decomposition~\cite{Hatano2005} as,
\begin{eqnarray}\label{Trotter}
e^{-\Delta\tau H}&=&e^{-\Delta\tau \sum_{i=1}^mH_i}\\ \nonumber
&=&\prod_{i=1}^me^{-\Delta\tau H_{i}}\prod_{i=m}^1e^{-\Delta\tau H_{i}}+\mathcal O(\Delta\tau^2) \, .
\end{eqnarray}
Using Eq.~(\ref{eq:TE}), we have,
\begin{eqnarray}
\label{eq:TEV1}
e^{-\tau H}&=&\Big[\prod_{i=1}^me^{-\Delta\tau H_{i}}\prod_{i=m}^1e^{-\Delta\tau H_{i}}+\mathcal O(\Delta\tau^2)\Big]^N\\ \nonumber
&=&\Big[\prod_{i=1}^m\prod_je^{-\Delta\tau H_{i,j}}\prod_{i=m}^1\prod_je^{-\Delta\tau H_{i,j}}\Big]^N+\mathcal O(\Delta\tau) \, .
\end{eqnarray}
We have approximated the original global operator $e^{-\tau H}$ by a successive local actions of $e^{-i\Delta \tau H_{i,j}}$, with controllable precision by $\delta\tau$.

\begin{figure} [!hbp]
		\begin{center}
		\resizebox{90mm}{!}{\includegraphics{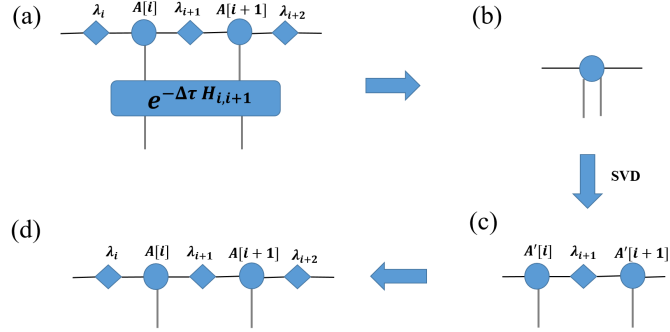}}
		\caption{The procedure of one step imaginary time evolution $e^{-\Delta\tau H_{i,j}}|\Psi_{\rm MPS}\rangle$\citep{Vidal2007}.
The diamonds are the environment matrices $\lambda_i$, whereas the circles are the tensors $A[i]$. The time evolution operator is a tensor with 4 legs.
We first contract the tensors $A[i]$, $A[i+1]$, $\lambda_i$, $\lambda_{i+1}$, $\lambda_{i+2}$ and ${\rm exp}\{-\Delta\tau H_{i,i+1}\}$ (a); We then perform
a SVD to the resulting 4-leg tensor (b); and obtain $A'[i]$, $A'[i+1]$ and $\lambda_i$ (c); Finally we restore the original format of the MPS,  as $A[i]$=$\lambda_i^{-1}A'[i]$ ,  $A^{[i+1]}$=$A'[i+1]\lambda_{i+2}^{-1}$ (d).   }\label{TEBD}
		\end{center}
\end{figure}

We now apply the imaginary time evolution Eq.~(\ref{eq:TEV1}) to the MPS\citep{Vidal2007}. For simplicity, we consider the local Hamiltonian $H_{i,i+1}$ only acts on two nearest neighbor $i$th and $(i+1)$th sites. We show graphically how $e^{-\Delta\tau H_{i,i+1}}$ acts on an MPS state in Fig.~\ref{TEBD}(a-d).
For convenience, we slightly modify the form of the MPS, by introducing additional diagonal matrices $\lambda_i$ ($i=1,2,\cdots, N-1$) into the MPS, which are the Schmidt coefficients of the MPS when it is divided into two parts at site $i$. $\lambda_i$  is also known as the ``environment''.
For each site, there is a tensor $A[i]$ represented by a circle, and a matrix $\lambda_i$ represented by a diamond.
The time evolution operation $e^{-\Delta\tau H_{i,i+1}}$ can be expressed as a tensor with four physical legs.
The whole processes for one step time evolution  are shown in Fig.~\ref{TEBD}.

First, we contract the tensors $A[i]$, $A[i+1]$, $\lambda_i$, $\lambda_{i+1}$, $\lambda_{i+2}$ and ${\rm exp}\{-\Delta\tau H_{i,i+1}\}$ [Fig.~\ref{TEBD}(a)],
and the resulting tensor is shown in Fig.\ref{TEBD}(b).
After the action of the  ${\rm exp}\{-\Delta\tau H_{i,i+1}\}$, the original format of MPS is modified, where $A[i]$, $A[i+1]$ fuse to a new tensor with four legs.
We perform the singular value decomposition (SVD) to separate site $i$ and $i+1$.
The left and right SVD tensors are temporarily labeled by $A'[i]$, $A'[i+1]$ respectively, and the matrix $\lambda_{i+1}$.
The virtual dimension between the site $i$ and $i+1$ will be $D\times d$ where $d$ is the dimension of the physical index.
In order to find an optimal approximation of the modified MPS with fixed parameter $D$,
we only keep the $D$ largest singular values in $\lambda_{i+1}$ and their corresponding vectors $A'[i]$, $A'[i+1]$ [Fig.~\ref{TEBD}(c)].
Finally, in order to restore the original format of MPS  at site $i$ and $i+1$, we need to take off the effect of the environment and update
$A[i]$=$\lambda_i^{-1}A'[i]$ and $A^{[i+1]}$=$A'[i+1]\lambda_{i+2}^{-1}$ [Fig.~\ref{TEBD}(d)].

With a set of local imaginary time evolution operators according to Trotter decomposition of a given Hamiltonian $H$, we can efficiently find the optimal ground state of $H$ in the MPS space with fixed $D$, which will converge to the real ground state of $H$ with the increasing of $D$.

\subsubsection{Variational method}

Besides the imaginary time evolution method, we can also optimize an MPS using a variational method, i.e., to minimize the energy $\langle\Psi|H|\Psi\rangle$, with the constrain $\langle \Psi|\Psi\rangle=1$.
The application of Lagrange multiplier method transforms the question into minimizing the quantity
\begin{equation}\label{mini_Lagrange}
	\langle\Psi|H|\Psi\rangle-\lambda(\langle\Psi|\Psi\rangle-1)\, ,
\end{equation}
with respect to both unnormalized $|\Psi\rangle$ and the multiplier $\lambda$.

\begin{figure} [!hbp]
		\begin{center}
		\resizebox{100mm}{!}{\includegraphics{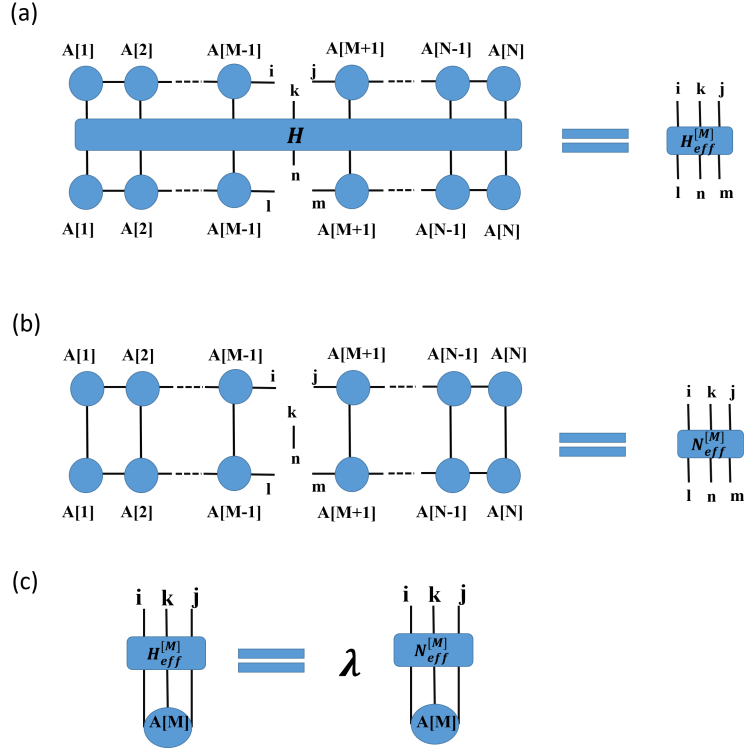}}
		\caption{The tensor network for (a) $H_{\rm eff}^{[M]}$ and (b) $N_{\rm eff}^{[M]}$, where the circles are the tensors $A[i]$,
  and the square represents the Hamiltonian $H$.
The resulting $H_{eff}^{[M]}$ and $N_{\rm eff}^{[M]}$ are 6-leg tensors, where $i$, $j$, $k$, $l$, $m$, $n$ are uncontracted indices.(c) Tensor network representation of equation $ H_{\rm eff}^{[M]}A[M]=\lambda N_{\rm eff}^{[M]} A[M] $, where the circle is the eigenvectors $A[M]$, and squares represents the effective Hamiltonian $H_{\rm eff}^{[M]}$ and $N_{\rm eff}^{[M]}$
}\label{eigen}
		\end{center}
\end{figure}

The stationary solution of Eq.~(\ref{mini_Lagrange}) by taking derivatives with respect to $\lambda$ and tensor elements, leads to the equations\cite{Schollwoeck2012}
\begin{equation}\label{mini_Lagrange2}
\left\{\begin{array}{l}
H_{\rm eff}^{[M]}A[M]=\lambda N_{\rm eff}^{[M]} A[M]\\
\langle\Psi|\Psi\rangle=1
\end{array}
\right.
\end{equation}
where $A[M]$ is the $M^{th}$ tensor of the MPS, and $H_{\rm eff}^{[M]}$ and $N_{\rm eff}^{[M]}$ are tensors defined in Fig.~\ref{eigen}. Since the second equation can always be satisfied by rescaling the state, we only need to care about the first equation, which can be expressed graphically in Fig.~\ref{eigen} (c).
This equation is actually a generalized eigenvalue problem for each $M$ and can be easily solved for a tensor at one time. \cite{Schollwoeck2012}
In practice, we solve the equation for $M$ sequentially from $1$ to $L$ and back to $1$. We repeat the process
until the result converges. With this method, we can find the optimal MPS with fixed $D$ to approximate the ground state of $H$. By increasing $D$, the
MPS may converge to the real ground state.

\subsection{Calculation of the expectation value of physical observables}

\begin{figure} [!hbp]
		\begin{center}
		\resizebox{60mm}{!}{\includegraphics{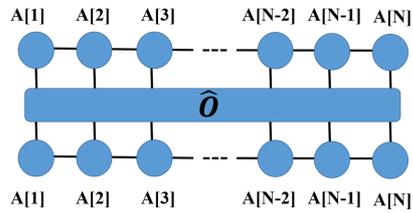}}
		\caption{The tensor network for $\langle O \rangle =\langle\Psi_{\rm MPS}|O|\Psi_{\rm MPS}\rangle$, where the circles are the tensors $A[i]$.
}\label{Value}
		\end{center}
\end{figure}

 After we obtain the ground state represented in MPS, we can calculate the physical observables to investigate the properties of the state.
 To obtain the expectation value of an operator, $\langle O\rangle=\langle\Psi|\hat{O}|\Psi\rangle$,
 we need to contract the tensor network shown in Fig.~\ref{Value}, where the operator $O$ is treated as a tensor.
 In practice, we usually express $\hat{O}$ as sum of local terms and calculate them separately.\cite{Schollwoeck2012}

\subsection{Other applications of TNS}
	
TNS can also be used to simulate dynamical systems and systems at finite temperature. Interestingly, it has been reported that MERA is an efficient expression of the conformal field on edge of an anti-deSitter(AdS) space\cite{Swingle12}, which is the holographic duality to the bulk space. This sheds light on our understanding of the nature of quantum gravity\cite{Susskind16}. Furthermore, making use of the resemblance of big data processing and statistical physics, TNS has been used in the fields of machine learning\cite{Cichocki14}.
Such applications can be conducted more efficiently using our tensor package.

\section{Basic operations in tensor network state methods}
\label{sec:operations}
	
As discussed in previous sections, there are different types of tensors that are used in different TNS methods, and there are also various methods to achieve the ground states and calculate expectation values. Therefore it is hard to develop a universal code for the TNS methods as those widely used first-principles packages
based on density functional theory. However, after careful analysis of the TNS methods, we find the TNS methods do share many similar basic operations, especially on the tensors. In this section, we summarize the commonly used basic operations in the TNS methods.

\subsection{Contraction of two tensors}

\begin{figure} [!hbp]
		\begin{center}
		\resizebox{90mm}{!}{\includegraphics{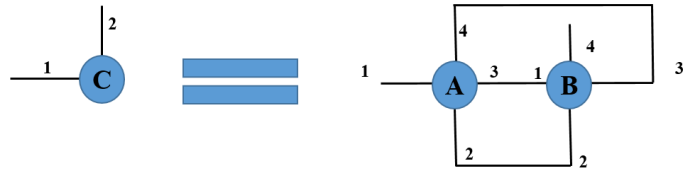}}
		\caption{The contraction of tensors $A$ and $B$ results in tensor $C$.
The numbers label the order of the legs in the tensors.}\label{contractpic}
		\end{center}
\end{figure}

Contraction of two tensors is one of the key operations that are widely used in TNS.
Generally, when two tensors are connected with legs, we need to sum the bond dimension of these legs. This process is called ``contraction''.
As an example, the contraction of two tensors $A_{j_1,\alpha_1,\alpha_2,\alpha_3}$ and $B_{\alpha_2,\alpha_1,\alpha_3,j_2}$ are schematically
shown in Fig.~\ref{contractpic}. It tells us that we should sum all connected bonds, namely the 2nd leg of $A$ with 2nd leg of $B$, the 3rd leg of $A$ with the 1st leg of $B$ and the 4th leg of $A$ with the 3rd leg of $B$. The resulting tensor after the contraction
is a two-leg tensor $C$, with one leg comes from the first leg of $A$ and the other from the 4th leg of $B$.
The formula of the  graphical notation is:
\begin{equation}\label{contract}
C_{j_1,j_2}=\sum_{\alpha_1,\alpha_2,\alpha_3}A_{j_1,\alpha_1,\alpha_2,\alpha_3}B_{\alpha_2,\alpha_1,\alpha_3,j_2}\, .
\end{equation}
For spins and bosons, the order of the contraction is not important. However, for fermions, one must be careful about the contraction order.

Tensor contraction is one of the most time-consuming parts of the TNS method, especially when the tensors has many legs. One also has to be very careful to contact the correct legs.

\subsection{Fuse legs of a tensor}

\begin{figure} [!hbp]
		\begin{center}
		\resizebox{90mm}{!}{\includegraphics{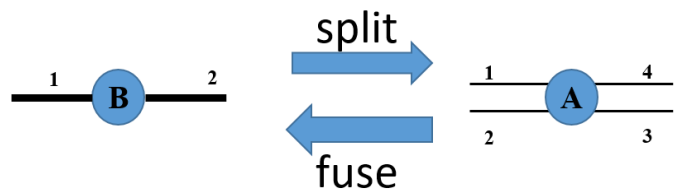}}
		\caption{The fusion of the first leg with the second leg, and the third leg and the fourth leg of $A$ results in a two-leg tensor $B$. Splitting is a reverse operation of fusion.}\label{fuse}
		\end{center}
\end{figure}

Sometimes we need to reshape a tensor, for example to a matrix to do singular value decomposition (SVD) etc. In this case, we need to combine
two or more legs of the tensor into a new leg, and the dimension of the new leg is the production of the dimensions of the original legs.
This process is called fusion of the legs, which are graphically shown in Fig.\ref{fuse}, where we fuse the 1st leg with the 2nd leg and the 3rd leg with the 4th leg of tensor $A$. After fusion,
the resulting tensor $B$ is a two-leg matrix.

\subsection{Split a leg of a Tensor}

The splitting of a leg of a tensor to two legs can be viewed as the inverse operation of the fusion operation. Generally, after a fusion operation and some other operations, we need a splitting operation to recover the original form of the tensor. Taking the example of Fig.~\ref{fuse},
we can split the two legs of $B$ back to four legs  in $A$. The order of legs in tensor $A$ should be carefully arranged to avoid confusion in future operations.

\subsection{Tensor leg permutation}

The {\it permutation} is used to reorder the legs of a tensor. It plays the key role in the fusion operation, because to fuse two legs, one need to permute the two legs to the adjacent positions.
The permutation operation can be very time-consuming for high rank tensors.

\section{Overview of the package}
\label{sec:overview}

The TNSpackage is written in Fortran2003. We use an object oriented programming (OOP) style to improve the readability and re-usability. So far it has about 42000 lines, and more than 200 subroutines and functions which are grouped into 8 modules: Tensor.f90, function.f90, Dimension.f90, print\_in\_TData.f90, element\_in\_TData.f90, modify\_in\_TData.f90, permutation\_in\_TData.f90, TData.f90, among which Tensor.f90 is the main module. To use the package, one need to include the Tensor.f90 module in the codes.
The package have been tested to be a stable version. With high reusability, the users are able to write their own children objects through inheritance. The procedure polymorphism and data polymorphism of the package make it suitable for coding different algorithms for TNS or for other purposes involving high dimension arrays.

In the package, we define a new data type {\tt Tensor}, whose elements are stored in a one-dimensional array.
The new data type {\tt Tensor} supports all data types of Fortran, including integer, single/double precision real number, single/double precision complex number,
logical and character. We overwrite most functions of Fortran2003, such as dcmplx, max, '.eq.', '+', '-', '*', '/' and so on, so they
can be directly applied to {\tt Tensor}, as they are applied to other fundamental data types.

In Sec. 4, we summarize the most often used tensor operations in the TNS methods.
We design accordingly the functions/subroutines to simplify these tensor operations.
We list in Table \ref{tab:functions} the most important and often used functions and subroutines for TNS.

\begin{table} [h]
\begin{center}
\caption{Most used functions and subroutines in the  package.}
\begin{tabular}{ |c |  c| c|}
\hline
Functions & purpose &examples\\
/subroutines &  & \\
\hline
\hline
setName & give names to the dimension of the Tensors& call T\%setName(1,'T.L') \\
\hline
fuse & fuse 2 or more legs of a Tensor into one leg& call T\%fuse(1,3)\\
\hline
split& reverse  operation of fuse& call T\%split()\\
\hline
permute & reorder the legs of a Tensor& call T\%permute([3,2,1])\\
\hline
contract & contraction of two Tensors &A=contract(B,1,C,2)\\
\hline
SVDTensor & SVD function&SVD=T\%SVDTensor() \\
\hline
eig & output the eigenvalue of the Tensor& eig=T\%eig() \\
\hline
QRTensor & QR decomposition of a Tensor &QR=T\%QRTensor\\
\hline
LQTensor & LQ decomposition of a Tensor &LQ=T\%LQTensor \\
\hline
max & output the max element of the Tensor& a=T\%max() \\
\hline
min & output the min element of the Tensor&a=T\%min() \\
\hline
norm & output the norm of the Tensor&a=T\%norm() \\
\hline
pointer & output a pointer which point to the data of the Tensor&call T\%pointer(data) \\
\hline
\end{tabular}
\end{center}
\label{tab:functions}
 \end{table}

We carefully design the data structure of {\tt Tensor} to speed up the performance and make our package more user-friendly. Some high rank tensor operations can be very time consuming if not designed wisely. For example, reshaping tensors (including fuse, and split the legs etc.)
may be time-consuming if one really move the data around in the memory, but in our package the operations of fusing or splitting legs of a tensor will cost almost no CPU time as we only modify data structure of the tensor, which does not actually move the data themselves.
We also make great effort to optimize the {\it permeation} operation which is heavily used in the code.

We use high performance mathematic packages including Arpack \cite{Arpack}, Lapack \cite{LapackAndBlas} and Blas \cite{LapackAndBlas}
to improve the efficiency for linear algebraic operations, such as contractions (matrix multiplications), QR decompositions, SVD etc..
Direct use these subroutines are tedious and cumbersome.
To ease the coding work, these mathematic subroutines have been encapsulated in the TNSpackage,
and the users who use these packages through calling subroutines of TNS package, do not need to worry about the details on how they are implemented.
For example, the contraction of two tensors (See Eq.\ref{contract}), involves the product of two matrices or a matrix to a vector.
The TNSpackage calls XGEMM for the product of two matrices and XGEMV for the product for a matrix and a vector, where X is 'S', 'D', 'C', 'Z' standing for single/double real/complex data. The user can just use a short code like ``{\tt A=contract(B,1,C,2)}'', to contract the first leg of B with the second
leg of C, and the package will automatically choose proper subroutines and parameters for the operations,
which greatly simplifies the coding work. Similarly, one can do SVD, QR/LQ decompositions and other linear algebraic operations
in this simple way.

As shown in Sec. 4, most of the basic operations of tensors are on some specified legs of the tensors. In TNS, each leg has its physical meaning, and the order of the legs is of great importance. However, in many operations, the tensor legs are permuted, fused or split.
As a result, it will be very difficult to trace the correct legs after a few operations on the tensors.
One has to be extremely careful to ensure the operation is on the correct legs.
To overcome this complexity in coding TNS, we may assign a 'name' to every leg of the Tensor. The 'name' of the tensor legs are characters in the form of '{\tt TensorName.DimensionName}', where {\tt TensorName} is used to track the tensor name which the legs belong to and {\tt dimensionName} is to track the legs
in the tensor. We use "." to divide the tensor name and leg name. In the process of contraction, the legs that have been contracted (including their names) will be removed, but the remaining ones are saved in the new {\tt Tensor}. By reading the name of the leg we can easily tell which {\tt Tensor} the leg original
comes from and its physical meaning. All the functions/subroutines in Table I can be used by specifying legs with {\tt TensorName.DimensionName}.
With the help of the {\tt TensorName.DimensionName}, one do not need to consider
the actual order of the legs in the tensors, which greatly simplifies the coding work.

\section{Example codes for PEPS}
\label{sec:examples}

In this section, we demonstrate how to use the TNSpackage by an example of solving the Heisenberg model on
a square lattice using PEPS. The Hamiltonian reads,
\begin{equation}
H=\sum_{\langle i,j\rangle} {\bf S}_i \cdot {\bf S}_j \, ,
\end{equation}
where $\sum_{\langle i,j\rangle}$ denote the summation of the nearest-neighbor pairs. We assume the periodic boundary condition.
The wave functions of the Heisenberg model are presented PEPS, and we optimize the wave functions using a simple update time evolution method\cite{Xiang2008}.
We first give examples on some important ingredients of this algorithm to introduce step by step
the usage of the TNS package. The full program is given in the Appendix.

\begin{figure} [!hbp]
		\begin{center}
		\resizebox{100mm}{!}{\includegraphics{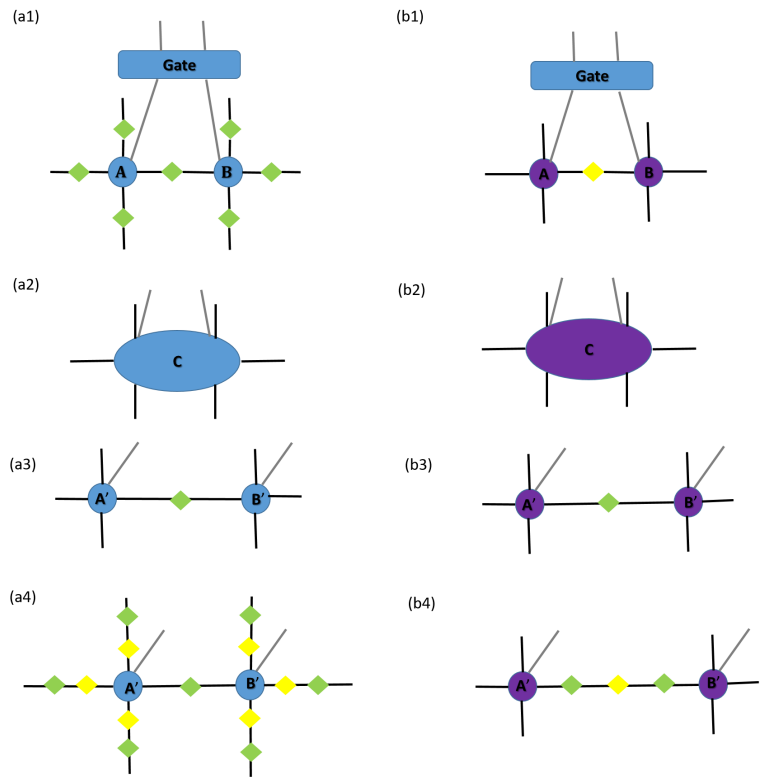}}
		\caption{The green diamonds denote the environment tensors and the yellow ones are the inverse of them.
The circles are the node in the tensor network. Left panel: The standard procedure of one step imaginary time evolution.
(a1) Contracting tensors {\tt A}, {\tt B}, with all environments and the time
evolution gate. (a2) The resulting tensor {\tt C} after contraction.
(a3) The resulting tensors obtained by performing SVD on the tensor in (a2). (a4) Restore the original form of tensors in (a1) by
contracting the inverse of the environments (the yellow diamonds).  Right panel: The implemented scheme of one step imaginary time evolution.
The stored tensors {\tt A} and {\tt B} have already included the environments. (b1)
Contract tensor {\tt A}, {\tt B}, the inverse of the environment between them and the time evolution gate.
(b2) The result of (b1) which is equivalent to (a2). (b3) The resulting tensors obtained by performing SVD on the tensor in (b2), which
is equivalent to (a3). (b4) Contract the  new environment (the green ones)
and restore the original tensor form in (b1).
This scheme may save three environment contractions from the standard procedures.}\label{fig::evo}
		\end{center}
\end{figure}

\subsection{Data structure of PEPS wave functions: }
We first design a data structure {\tt type Node}, to store all necessary data on each lattice site.
\begin{verbatim}
   type Node
      type(Tensor)::site  ! to store PEPS tensor of each site.
      type(Tensor)::Up    ! to store the tensor of the environment
      type(Tensor)::Down  ! to store the tensor of the environment
      type(Tensor)::Left  ! to store the tensor of the environment
      type(Tensor)::Right ! to store the tensor of the environment
   end type Node
\end{verbatim}
{\tt Tensor site} is to store the main data
of PEPS wave functions, which is a rank 5 tensor, including 4 ranks for the 4 virtual bonds, and one to store the physical indices.
Besides the main PEPS tensor, we also define 4 tensors: {\tt Up}, {\tt Down}, {\tt Left} and {\tt Right}
to store the environment information for the 4 virtual bonds, each of them is a rank 2 tensor, i.e., a matrix.
The graphic representations of {\tt Node A, B}  are shown in Fig.~\ref{fig::evo}(a).

\subsection{Generate PEPS wave functions: }

In the following subroutine, we generate a random PEPS as a starting wave function on the L1$\times$L2 lattice.
{\tt A(i,j)} with {\tt i}=1, ..., {\tt L1}, and {\tt j}=1, ..., {\tt L2}, are an array of data type {\tt Node}.
We first initialize tensor {\tt A(i,j)\%site} in each {\tt Node} structure to store the PEPS wave function,
where the first 4 legs with dimension {\tt D} is to store the virtual bonds, and the
last on to store the physical index, with $d=2$ (spin up and down). We define the tensor elements as single precision real numbers by specifying the variable {\tt datatype} in the code.
For convenience, we assign a name to each leg of {\tt A(i,j)\%site} to avoid confusion in future operations.
Next we set up the environment tensors {\tt Up}, {\tt Down}, {\tt Left}, {\tt Right} as identity matrices.

\begin{verbatim}
   subroutine initialPEPS(A,L1,L2,D)
      type(Node),allocatable,intent(inout)::A(:,:)
      integer,intent(in)::L1,L2,D
      integer::i,j
      allocate(A(L1,L2))
      do j=1,L2
         do i=1,L1
1           A(i,j)%site=generate([D,D,D,D,2],datatype)
2           call A(i,j)%site%setName(1,'A'+i+'_'+j+'.Left')
            call A(i,j)%site%setName(2,'A'+i+'_'+j+'.Down')
            call A(i,j)%site%setName(3,'A'+i+'_'+j+'.Right')
            call A(i,j)%site%setName(4,'A'+i+'_'+j+'.Up')
            call A(i,j)%site%setName(5,'A'+i+'_'+j+'.phy')
3           call A(i,j)%Up%allocate([D,D],datatype)
4           call A(i,j)%Up%eye()
            call A(i,j)%Up%setName(1,'Lambda.Leg1')
            call A(i,j)%Up%setName(2,'Lambda.Leg2')
5           A(i,j)%Down=A(i,j)%Up
            A(i,j)%Left=A(i,j)%Up
            A(i,j)%Right=A(i,j)%Up
         end do
      end do
      return
   end subroutine

\end{verbatim}
where
\begin{enumerate}
	\item[1] Generate {\tt site} as a ${\tt D}\times {\tt D}\times {\tt D}\times {\tt D}\times 2$ Tensor, whose elements are random real numbers. The data type is which are specified by the parameter {\tt datatype}.
	\item[2] Set the name {\tt Ai\_j\%Left} to the first leg the the Tensor.  Here {\tt i} and {\tt j} are integers, which are used for the lattice site indices. We have overwrite the operator ({\tt +}), so {\tt i} and {\tt j} are treated as characters in the operation {\tt 'A'+i+'\_'+j+'.Left'}.
%	\item[3] Set the name 'Ai\_j.Down' to the second leg the the Tensor.
%	\item[4] Set the name 'Ai\_j.Right' to the third leg the the Tensor.
%	\item[5] Set the name 'Ai\_j.Up' to the forth leg the the Tensor.
%	\item[6] Set the name 'Ai\_j.phy' to the fifth leg the the Tensor.
	\item[3] Allocate memory for {\tt A(i,j)\%Up}, which is a  ${\tt D} \times {\tt D}$ matrix, and the data type is real number.
	\item[4] Set the {\tt A(i,j)\%Up} as a identity matrix.
    \item[5] Copy tensor {\tt A(i,j)\%Up} to {\tt A(i,j)\%Down}.
\end{enumerate}
%When working on the open boundary conditions, the virtual bond dimensions on the boundary should be set to 1.

The data type of a tensor can be integer, real(kind=4), real(kind=8), complex(kind=4), complex(kind=8),
logical and character as listed in Table ~\ref{tab:datatype}.
\begin{table} [h]
\begin{center}
\caption{"Datatype" and its value}
\begin{tabular}{ |c |  c|}
\hline
datatype & data type of Tensor \\
\hline
\hline
'integer' & integer \\
\hline
'real' & real(kind=4) \\
\hline
'real*4' & real(kind=4) \\
\hline
'real(kind=4)' & real(kind=4) \\
\hline
'double' & real(kind=8) \\
\hline
'real*8' & real(kind=8) \\
\hline
'real(kind=8)' & real(kind=8) \\
\hline
'complex' & complex(kind=4) \\
\hline
'complex*8' & complex(kind=4) \\
\hline
'complex(kind=4)' & complex(kind=4) \\
\hline
'complex*16' & complex(kind=8) \\
\hline
'complex(kind=8)' & complex(kind=8) \\
\hline
'logical' & logical \\
\hline
'character' & character(len=*) \\
\hline
\end{tabular}
\end{center}
\label{tab:datatype}
 \end{table}
User can use the following code to specify the data type of a Tensor:
\begin{verbatim}
   type(Tensor)::T
   character(len=20)::datatype
   datatype='complex*8'
   call T%setType(datatype)
\end{verbatim}
where the character {\tt datatype} specify the data type of {\tt T}.
If one do not specify the data type of the {\tt Tensor},
the package will automatically choose a suitable data type for it.
For example, we can copy a tensor to another using the following code:
\begin{verbatim}
   T1=T2 !copy tensor T2 to T1
\end{verbatim}
Here, if the data type of {\tt T1} have been set before by calling {\tt T1\%setType(datatype)},
the data type will be determined by the specified data type, otherwise, the data type is set to the same as that of {\tt T2}.

\subsection{Define the time evolution operator as a quantum gate:}

 To calculate the ground state of the model using the imaginary time evolution method, we need to define the Hamiltonian
 in terms of two-body operators $H_{1,2}={\bf S}_1\cdot {\bf S}_2=(S_1^x\otimes S_2^x)+(S_1^y\otimes S_2^y)+(S_1^z\otimes S_2^z)$, where $S_{1,2}^{x,y,z}$ are the $\frac{1}{2}$ Pauli matrix and $\otimes$ is the direct product operator.
 The operator $H_{1,2}$ can be viewed as a 4-leg tensor. In the following subroutine, we
 define a gate operator ${\rm Gate}={\rm exp}(-\tau H_{1,2})$, where $\tau$ is the length of the time step,
  to perform time evolution between two sites. We first reshape the 4-leg tensor $H_{1,2}$ to a matrix(2-leg Tensor) by fusing leg 1 with leg 2, and leg 3 with leg 4. We then perform time evolution $e^{-\tau H_{1,2}}$ on the matrix. Finally, we reshape (split) the resulting matrix to its original form of a 4-leg tensor.

\begin{verbatim}

   subroutine initialH(H)
      type(Tensor),intent(inout)::H
      type(Tensor)::Sx,Sy,Sz
1     call pauli_matrix(Sx,Sy,Sz,0.5)
      call H%setType(datatype)
2     H=(Sx.xx.Sx)+(Sy.xx.Sy)+(Sz.xx.Sz)
3     call H%setName('H')
      return
   end subroutine

   type(Tensor) function gate(H,tau)
      type(Tensor),intent(in)::H
      real,intent(in)::tau
      gate=H*tau
4     call gate%fuse(1,2)
      call gate%fuse(2,3)
5     gate=expm(gate)
6     call gate%split()
      return
   end function
\end{verbatim}
where
\begin{enumerate}
	\item[1] Generate the 1/2 pauli matrices for $S_x$, $S_y$ and $S_z$.
	\item[2] The operator ({\tt .xx.}) is the direct product operator, e.g., {\tt C=A.xx.B} is defined as $C_{i,j,k,l}=A_{i,k}*B_{j,l}$.
	\item[3] Set TensorName '{\tt H}' to tensor {\tt H}, and the legs of tensor {\tt H} are automatically named by order: '{\tt H.1}', '{\tt H.2}',
'{\tt H.3}' and '{\tt H.4}'.
	\item[4] Fuse the 1st and the 2nd legs of tensor '{\tt gate}' into a new leg.
	\item[5] Calculate $e^{A}$, where $A$ is a matrix
	\item[6] Split all the legs of the '{\tt gate}' tensor back to it original form (a 4 leg tensor).
\end{enumerate}

\subsection{Imaginary time evolution:}
\label{sec:ITE}

 One step time evolution involves 3 steps as shown in Fig.~\ref{fig::evo} (a1) - (a4): (i) Contract the environments to the tensors {\tt A}, {\tt B}, and then contract  with gate. The resulting tensor {\tt C} is shown in Fig.~\ref{fig::evo}(a2). (ii) Do SVD to tensor {\tt C} [Fig.~\ref{fig::evo}(a3)];
(iii) Contract the inverse of the environments, so all tensors are back to theirs original formats [Fig.~\ref{fig::evo}(a4)].
In our implementation, we slightly change the procedures, which are shown
in Fig.~\ref{fig::evo}(b1) - (b4). In our scheme, tensors {\tt A} and {\tt B} are the ones that already have contracted the environments.
Therefore at the first step, one should contract the inverse of the environment between the tensor A and the tensor B.
 At the final step, both tensor {\tt A} and tensor {\tt B} need to contract the new environment tensor.
In this scheme, one can save three environment contractions in each time step from the standard procedures.

\begin{figure} [!hbp]
		\begin{center}
		\resizebox{100mm}{!}{\includegraphics{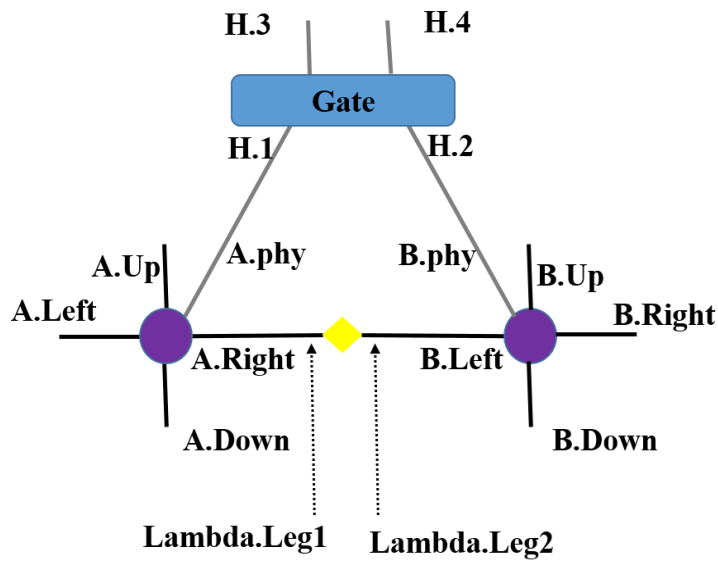}}
		\caption{The contraction of tensors {\tt A}, {\tt B} with the inverse of environment and the time evolution gate.
The yellow diamond stands for the inverse of the environment, whereas the purple circles are the tensors, with the names of the legs. The physical legs
are in grey.}
\label{fig::evo2}
		\end{center}
\end{figure}

In the following subroutine, we demonstrate how to use the gate tensor to perform time evolution of two sites {\tt A} and {\tt B}.
For convenience, we assign names to the legs of the two tensors and the gate, where
{\tt AName} is the {\tt TensorName} of tensor {\tt A} and {\tt BName} is the {\tt TensorName} of {\tt B}; {\tt cha1} and {\tt cha2} specify the legs to be contracted
during time evolution. {\tt Lambda} is the environment between {\tt A} and {\tt B}.
The contraction of time evolution gate is shown in Fig.~\ref{fig::evo2}, where {\tt AName}='{\tt A}', {\tt BName}='{\tt B}',
{\tt cha1}='{\tt Right}' and {\tt cha2}='{\tt Left}'. The yellow diamond is the inverse of the environment ({\tt Lambda}$^{-1}$) between tensors {\tt A} and {\tt B}.

\begin{figure} [!hbp]
		\begin{center}
		\resizebox{140mm}{!}{\includegraphics{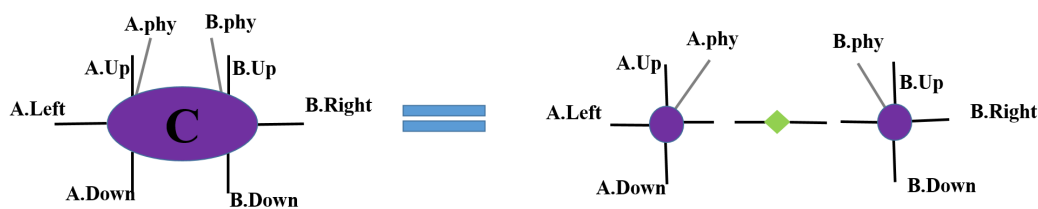}}
		\caption{ The SVD of tensor {\tt C} after time evolution gate operation (Fig.~\ref{fig::evo2}). The green diamond stands for the environment
{\tt Lambda}. The new legs come out from site {\tt A} and site {\tt B} do not have names, where other legs keep their original names before gate operation.
}\label{fig::SVD}
		\end{center}
\end{figure}

\begin{verbatim}
   subroutine step(A,Lambda,B,expmH,CutOff,AName,BName,cha1,cha2)
      type(Tensor),intent(inout)::A,Lambda,B
      type(Tensor),intent(in)::expmH
      character(len=*),intent(in)::cha1,cha2,AName,BName
      integer,intent(in)::CutOff
      type(Tensor)::C,SVD(3)
1     C=contract(A,AName+'.'+cha1,Lambda%invTensor(),'Lambda.Leg1')
2     C=contract(C,'Lambda.Leg2',B,BName+'.'+cha2)
3     C=contract(C,[AName+'.phy',BName+'.phy'],expmH,['H.1','H.2'])
4     call C%setName('H.3',AName+'.phy')
5     call C%setName('H.4',BName+'.phy')
6     SVD=C%SVDTensor(AName,BName,CutOff)
7     Lambda=eye(SVD(2))/SVD(2)%smax()
8     A=SVD(1)*Lambda
9     call A%setName(A%getRank(),AName+'.'+cha1)
      B=Lambda*SVD(3)
10    call B%setName(1,BName+'.'+cha2)
      call Lambda%setName(1,'Lambda.Leg1')
      call Lambda%setName(2,'Lambda.Leg2')
      return
   end subroutine
\end{verbatim}
where
\begin{enumerate}
	\item[1] Contract leg {\tt AName.cha1} from tensor {\tt A} with leg {\tt Lambda.Leg1} from environment tensor {\tt invTensor(Lambda)}, where
function {\tt invTensor()} gives the inverse of a matrix.
	\item[2] Contract leg  {\tt Lambda.Leg2} from tensor {\tt C} with leg {\tt BName.cha2} from tensor {\tt B}.
	\item[3] Contract leg {\tt AName.phy} (spin index of tensor {\tt A}) and leg {\tt BName.phy} (spin index of tensor {\tt B}) from tensor {\tt C} with the legs {\tt H.1} and {\tt H.2} from gate tensor {\tt expmH}.
	\item[4] Rename leg {\tt H.3} of tensor {\tt C} to {\tt AName.phy}.
	\item[5] Reanme leg {\tt H.4} of tensor {\tt C} to {\tt BName.phy}.
	\item[6] Do {\tt SVD} to tensor {\tt C}, such that {\tt C=SVD(1)*eye(SVD(2))*SVD(3)}.
%, where the operator(*) means contracting the last leg of the first tensor with the first leg of the second one and the
Function {\tt eye()}  reformats a vector to a diagonal matrix. The legs from {\tt AName} are regarded as row, and the legs from {\tt BName} are regarded as col. In output, {\tt SVD(1)} are the left singular vectors, {\tt SVD(2)} store the singular values and {\tt SVD(3)} are the right vectors.
After SVD, the resulting two tensors automatically restore the structure of tensor {\tt A} and {\tt B}, according to tensor names (see Fig.\ref{fig::SVD})
, except the last leg of tensor {\tt A} and first leg of tensor {\tt B}, whose dimensions are
determined by the input parameters {\tt cutoff}. This procedure is shown in Fig.~\ref{fig::SVD}.
    \item[7] Update environment tensor {\tt Lambda}, which is normalized by its maximum element.
    \item[8] Update PEPS tensor {\tt A} after time evolution.
    %As the schedule of our evolution, A store the tensor who is connected to all its environments.}
    \item[9] Set name {\tt AName.+cha1} to the last leg of the updated tensor {\tt A}, which has been
    contracted during the time evolution, so it can be used in the next time evolution steps. {\tt A\%getRank()} get the rank (number of legs) of tensor {\tt A}.
    \item[10 ] Set name {\tt BName.+cha2} to the first leg of the updated tensor {\tt B}, which has been
    contracted during the time evolution.
\end{enumerate}

If one don't want to use the leg names for the operations, one may also explicitly specify the legs by their orders in the tenors
to be used. However, doing this way will be more complicated and elusive than using tensor and leg names,
because one has to track the orders of tensors' legs which may change during the operations.

\section{Summary}
\label{sec:summary}

We have developed a Fortran2003 package namely TNSpackage that integrates most often used tensor operations
designed for the TNS methods. Great efforts has been spent to optimize the efficiency of the tensor operations.
The package is user-friendly and flexible for different types of TNS, and
therefore greatly simplifies the coding work for developing the TNS methods. The package is also very useful for
other algorithms that involving high rank tensor operations.

\section*{Acknowledgments}
The authors thank Prof. Hong An and Chao Yang for valuable discussions.
This work was funded by the Chinese National Science Foundation (Grant number 11374275,11474267),
the National Key Research and Development Program of China (Grants No. 2016YFB0201202).
The numerical calculations have been done on the USTC HPC facilities.

\section*{Appendix}

In this appendix, we give a complete example of imaginary time evolution of the Heisenberg model on a 4$\times$4 square lattice, with
the periodic boundary condition.
We use simple update method, proposed in Ref.~\cite{Xiang2008}.
\begin{verbatim}
module example_PEPS
   use Tensor_type
   use usefull_function
   implicit none
   character(len=20)::datatype='real'
   type Node
      type(Tensor)::site
      type(Tensor)::Up
      type(Tensor)::Down
      type(Tensor)::Left
      type(Tensor)::Right
   end type Node
contains

   subroutine initialH(H)
      type(Tensor),intent(inout)::H
      type(Tensor)::Sx,Sy,Sz
      call pauli_matrix(Sx,Sy,Sz,0.5)
      call H%setType(datatype)
      H=(Sx.xx.Sx)+(Sy.xx.Sy)+(Sz.xx.Sz)
      call H%setName('H')
      return
   end subroutine

   type(Tensor) function gate(H,tau)
      type(Tensor),intent(in)::H
      real,intent(in)::tau
      gate=H*tau
      call gate%fuse(1,2)
      call gate%fuse(2,3)
      gate=expm(gate)
      call gate%split()
      return
   end function

   subroutine initialPEPS(A,L1,L2,D)
      type(Node),allocatable,intent(inout)::A(:,:)
      integer,intent(in)::L1,L2,D
      integer::i,j
      allocate(A(L1,L2))
      do j=1,L2
         do i=1,L1
            A(i,j)%site=generate([D,D,D,D,2],datatype)
            call A(i,j)%site%setName(1,'A'+i+'_'+j+'.Left')
            call A(i,j)%site%setName(2,'A'+i+'_'+j+'.Down')
            call A(i,j)%site%setName(3,'A'+i+'_'+j+'.Right')
            call A(i,j)%site%setName(4,'A'+i+'_'+j+'.Up')
            call A(i,j)%site%setName(5,'A'+i+'_'+j+'.phy')
            call A(i,j)%Up%allocate([D,D],datatype)
            call A(i,j)%Up%eye()
            call A(i,j)%Up%setName(1,'Lambda.Leg1')
            call A(i,j)%Up%setName(2,'Lambda.Leg2')
            A(i,j)%Down=A(i,j)%Up
            A(i,j)%Left=A(i,j)%Up
            A(i,j)%Right=A(i,j)%Up
         end do
      end do
      return
   end subroutine

   subroutine step(A,Lambda,B,expmH,CutOff,AName,BName,cha1,cha2)
      type(Tensor),intent(inout)::A,Lambda,B
      type(Tensor),intent(in)::expmH
      character(len=*),intent(in)::cha1,cha2,AName,BName
      integer,intent(in)::CutOff
      type(Tensor)::C,SVD(3)
      C=contract(A,AName+'.'+cha1,Lambda%invTensor(),'Lambda.Leg1')
      C=contract(C,'Lambda.Leg2',B,BName+'.'+cha2)
      C=contract(C,[AName+'.phy',BName+'.phy'],expmH,['H.1','H.2'])
      call C%setName('H.3',AName+'.phy')
      call C%setName('H.4',BName+'.phy')
      SVD=C%SVDTensor(AName,BName,CutOff)
      Lambda=eye(SVD(2))/SVD(2)%smax()
      A=SVD(1)*Lambda
      call A%setName(A%getRank(),AName+'.'+cha1)
      B=Lambda*SVD(3)
      call B%setName(1,BName+'.'+cha2)
      call Lambda%setName(1,'Lambda.Leg1')
      call Lambda%setName(2,'Lambda.Leg2')
      return
   end subroutine

   subroutine sampleUpdate(A,expmH,L1,L2,CutOff)
      type(Node),intent(inout)::A(:,:)
      type(Tensor),intent(in)::expmH
      integer,intent(in)::L1,L2,CutOff
      integer::i,j
      do i=1,L1
         do j=1,L2-1
            call step(A(i,j)%Site,A(i,j)%Right,A(i,j+1)%Site&
              ,expmH,CutOff,'A'+i+'_'+j,'A'+i+'_'+(j+1),'Right','Left')
            A(i,j+1)%Left=A(i,j)%Right
         end do
         call step(A(i,L2)%Site,A(i,L2)%Right,A(i,1)%Site&
              ,expmH,CutOff,'A'+i+'_'+L2,'A'+i+'_1','Right','Left')
         A(i,1)%Left=A(i,L2)%Right
      end do
      do j=1,L2
         do i=1,L1-1
            call step(A(i,j)%Site,A(i,j)%Down,A(i+1,j)%Site&
               ,expmH,CutOff,'A'+i+'_'+j,'A'+(i+1)+'_'+j,'Down','Up')
            A(i+1,j)%Up=A(i,j)%Down
         end do
         call step(A(L1,j)%Site,A(L1,j)%Down,A(1,j)%Site&
             ,expmH,CutOff,'A'+L1+'_'+j,'A1_'+j,'Down','Up')
         A(1,j)%Up=A(L1,j)%Down
      end do
      return
   end subroutine

   subroutine PEPSrun()
      integer::L1,L2,D,i,runningnum,runningTime
      real::tau
      type(Node),allocatable::A(:,:)
      type(Tensor)::H,expmH
      L1=4
      L2=4
      D=4
      tau=0.1
      runningnum=50
      call initialPEPS(A,L1,L2,D)
      call initialH(H)
      expmH=expmTensor(H,tau)
      do runningTime=1,4
         do i=1,runningnum
            call sampleUpdate(A,expmH,L1,L2,D)
            tau=tau*0.1
         end do
      end do
      return
   end subroutine
end module

program Example
   use example_PEPS
   call PEPSrun()
end
\end{verbatim}

\section*{References}
%\bibliographystyle{prsty}
%\bibliography{Mybib}

\end{document}